
\input harvmac

\overfullrule0pt
\def\half{\hbox{$1\over 2$}}
\def\third{\hbox{$1\over 3$}}
\def\R{{\bf R}}
\def\d{\partial}
\def\dd{\d\bar\d}
\def\w{\wedge}
\def\eps{\epsilon}
\def\det{{\rm det\,}}
\def\O{{\cal O}}
\def\o{{\scriptstyle\cal O}}
\def\Z{{\bf Z}}

\jot=5pt

\def\Title#1#2{\nopagenumbers\abstractfont\hsize=\hstitle\rightline{#1}%
\vskip.7in\centerline{\titlefont #2}\abstractfont\vskip .5in\pageno=0}

\Title{\vbox{\baselineskip12pt
\hbox{CLNS--92/1139}\hbox{IASSNS--HEP--92/14}}}
{Gauge Symmetries of the N=2 String}
\centerline{Amit Giveon\footnote{$^*$}{giveon@iassns.bitnet}}
\centerline{School of Natural Sciences}
\centerline{Institute for Advanced Study}
\centerline{Princeton, NJ\ \ 08540}
\vskip.1in
\centerline{Alfred Shapere\footnote{$^\dagger$}
{shapere@strange.tn.cornell.edu}}
\centerline{Newman Laboratory of Nuclear Science}
\centerline{Cornell University}
\centerline{Ithaca, NY\ \ 14853}

\vskip.2in
\noindent{\bf Abstract:}
We study the underlying gauge symmetry algebra of the $N=2$ string,
which is broken down to a subalgebra in any spacetime background.  For
given toroidal backgrounds, the unbroken gauge symmetries
(corresponding to holomorphic and antiholomorphic worldsheet currents)
generate area-preserving diffeomorphism algebras of null
2-tori. A minimal Lie algebraic closure containing all the gauge
symmetries that arise in this way, is the background--independent
volume--preserving diffeomorphism algebra of the target Narain torus
$T^{4,4}$.  The underlying symmetries act on the ground ring of
functions on $T^{4,4}$ as derivations, much as in the case of the
$d=2$ string.  A background--independent spacetime action valid for
noncompact metrics is presented, whose symmetries are
volume--preserving diffeomorphisms.  Possible extensions to $N=2$ and
$N=1$ heterotic strings are briefly discussed.

\Date{February 1992}

\newsec{Introduction: Gauge Symmetries of Strings}

String theories possess an infinite number of gauge symmetries.  In
any given spacetime background, most of these gauge symmetries will be
spontaneously broken.  This fact is fortunate for phenomenology's
sake, but makes it difficult to untangle the underlying symmetry
structure. Explicit knowledge of the full gauge algebra would be
useful in formulating background--independent string
field theories, in constructing effective actions, and perhaps in
understanding how the string selects a particular vacuum.  In
addition, it is not unreasonable to hope that gauge symmetries of this
sort could play just as vital a role in the second quantization of string
theories as they did for Yang--Mills theories.

It has been suggested \ref\W{E. Witten, Commun. Math. Phys. {\bf 118}
(1988) 411.} that there might be an unbroken phase of strings, perhaps
a topological field theory, in which the full gauge symmetry algebra
would be represented on the string's Hilbert space. Indeed, spacetime
backgrounds break spacetime diffeomorphism invariance, and a
background--independent formulation of string theory would presumably
be diffeomorphism--invariant, {\it i.e.}, topological. We thus expect
spacetime diffeomorphisms to play a crucial role in the full gauge
algebra.

In string theory, worldsheet symmetries are directly related to
symmetries of the spacetime effective action.  From the point of view
of the worldsheet, conformal field theories (CFTs) are associated with
classical string vacua \ref\rGSW{For a review, see M.B. Green, J.H.
Schwarz, and E. Witten, Superstring Theory (Cambridge: Cambridge Univ.
Press, 1987).}.  An exact symmetry algebra of a particular string
vacuum is generated by the operator product algebras of the
holomorphic $(1,0)$ currents and anti-holomorphic $(0,1)$ currents of
the CFT.  In general, a CFT can be deformed by truly marginal $(1,1)$
operators.  The action of the symmetry generators on such deformations
can be translated into an action on the couplings to the marginal
operators, and therefore gives rise to symmetries on the moduli space
of couplings.  (This procedure has been used in \ref\rGMR{M. Dine, P.
Huet and N. Seiberg, Nucl. Phys. {\bf B322} (1989) 301\semi A. Giveon,
N. Malkin and E. Rabinovici, Phys. Lett. {\bf B238} (1990) 57.} to
show that target space duality symmetries in the flat case are
residual discrete symmetries of this type, while restricting (locally)
to the physical moduli space.)  The couplings to $(1,1)$ operators
become massless fields in spacetime, and the symmetries of the CFT
thus translate into symmetries of the spacetime effective action.

The $(1,0)$ (or $(0,1)$) operators
which are not holomorphic (anti-holomorphic) in a given vacuum, may
become holomorphic (anti-holomorphic) at some other points of the
moduli space. In order to reveal the underlying symmetries of string
theory, one should, therefore, consider at least all such operators.
Such a program was discussed for low-energy effective actions of the
heterotic string \ref\rGP{A. Giveon and M. Porrati, Phys. Lett. {\bf
B246} (1990) 54\semi A. Giveon and M. Porrati, Nucl. Phys. {\bf B355}
(1991) 422. }, and an infinite--dimensional gauge algebra (called the
``duality--invariant string gauge algebra'') was introduced,
in order to construct an effective action valid throughout
the moduli space of toroidal compactifications.  However, the
understanding of stringy gauge symmetries thus gained was incomplete.
The algebra did not include gauge symmetries involving higher--spin
states.  Also, implementing it as a symmetry algebra of the effective
action required introducing unphysical ultramassive ghost fields.

The complete description of the gauge symmetries of the $N=1$
heterotic string is a difficult problem, due to the complexity of the
vertex operator algebra. However, there are simpler string theories,
for which one might hope the problem would be more tractable.

In this paper, we will investigate gauge symmetries of the closed
string with $N=2$ local worldsheet supersymmetry, in the critical
dimension \ref\rNTWO{M. Ademollo et.al., Phys.Lett.\ \bf 62B \rm (1976)
105 \semi J.D. Cohn, Nucl. Phys.\ \bf B284 \rm (1987) 349\semi M.
Corvi, V.A. Kosteleck\'y, and P. Moxhay, Phys. Rev.\ \bf D39 \rm
(1988) 1611\semi S. Mathur and S. Mukhi, Nucl.Phys.\ \bf B302 \rm
(1988) 130\semi N. Ohta and S. Osabe, Phys. Rev.\ \bf D39 \rm (1989)
1641.}\ref\rOV{H. Ooguri and C. Vafa, Nucl. Phys. {\bf B361} (1991)
469.}. As explained above, these may be regarded either as gauge
symmetries of the CFT in a particular background, or as symmetries of
the effective spacetime action. Both points of view --- worldsheet and
spacetime --- will prove useful.  On the worldsheet, we study the CFTs
associated to various backgrounds, and find gauge symmetry generators
within the algebra of on--shell vertex operators.  In section 2 we
discuss the $N=2$ string in Minkowski space, and in section 3 we
consider general toroidal backgrounds.  The gauge symmetries that we
find generate area--preserving diffeomorphisms of two--dimensional
null subspaces, and act on the ground ring of dimension $(0,0)$
operators as derivations.  We show that for any on--shell dimension
$(1,0)$ current, there is a toroidal background where the current
becomes holomorphic, and the gauge symmetry that it generates is
unbroken.

The gauge algebras in different backgrounds are all subalgebras of the
full underlying off--shell algebra we seek. A minimal Lie algebraic
closure, obtained by continuing the representation of the gauge
generators as derivations to off--shell momenta, is the
volume-preserving diffeomorphism algebra of the target Narain torus
\ref\rNAR{K.S. Narain, Phys. Lett.\ {\bf B169} (1986) 369.}.
(Other closures exist; one possibility is a lattice algebra
over the Narain lattice.)   The
underlying symmetries of the $N=2$ string have some remarkable
similarities with those of the $d=2$ string, as we describe at the
end of section 3.

An independent approach, from the point of view of spacetime, is to
look for a background--independent effective action that reproduces
the correlation functions of the $N=2$ string when expanded around a
particular background. This is the subject of section 4.  We will
succeed in finding such an action, valid for noncompact backgrounds,
whose off--shell symmetries are exactly those obtained from the
worldsheet operator algebra.  It is the minimal action reproducing all
on--shell amplitudes in the background $\R^{2,2}$.

The underlying symmetry algebra of the $N=2$ string may shed light on
the underlying structure of more realistic string theories.  In
section 5 we discuss a possible extension of our ideas to the
$N=2$ and $N=1$ heterotic strings.

\newsec{The $N=2$ String in Minkowski Space}
The critical $N=2$ string has perhaps the simplest vertex operator
algebra of any string theory. This is because, unlike other string
theories, it contains a finite number of physical degrees of freedom.

The worldsheet action of the $N=2$ string is \rOV
\eqn\eWA{
S_0=\int d^2z d^2\theta d^2\bar\theta\, K_0 (X,\bar X)
}
in terms of the $N=2$ chiral superfield
\eqn\eCS{
X^i(Z,\bar Z; \theta^-, {\bar\theta}^-)=x^i(Z,\bar Z)
+\psi^i_L(Z,\bar Z)\theta^-
+\psi^i_R(Z,\bar Z){\bar \theta}^-
+F^i(Z,\bar Z)\theta^-{\bar\theta}^-
$$
$$
Z=z-\theta^+\theta^-
}
(where $i=s,t$ denote complex spacelike and timelike components, and
bars denote complex conjugation).  For now, we will take as target
space $\R^{2,2}$, with $K_0(X, \bar X)= X_s \bar X_s - X_t \bar X_t$.

The $N=2$ string around flat space $\R^{2,2}$ has a single massless
degree of freedom $\phi$ describing K\"ahler deformations of the
background geometry.  The vertex operator to create a mode of $\phi$
with complex 2-momentum $p$ is
\eqn\eV{
V_p(X, \bar X )=\exp\, i (p\!\cdot\! \bar X +\bar p \!\cdot\! X)
}
One calculates tree--level correlation functions by inserting $V_p$ at
$n$ points on the sphere, and integrating over their positions modulo
global superconformal transformations.  Then one finds that the
on--shell three--point function is \rOV\
\eqn\eTHREE{
\langle V_p V_q V_r\rangle =
(q\!\cdot\! \bar r -r\!\cdot\! \bar q)^2
}
with $p\!\cdot\! \bar p= q\!\cdot\! \bar q = r\!\cdot\! \bar r=0$ and
$p+q+r=0$.  It turns out that four--point and probably also
higher--point amplitudes of such operators vanish.

The superfield vertex operator has the following expansion
in terms of component fields:
\eqn\eCOMP{
\eqalign{
V_p(X,\bar X)&=e^{i( p \cdot\bar x + \bar p\cdot x)}\cr
&\quad +
\left( ip\!\cdot\! \d \bar x - i \bar p
\!\cdot\! \d x - (p\!\cdot\! \bar\psi_L)(\bar p \!\cdot\! \psi_L)\right)
e^{i( p \cdot\bar x + \bar p\cdot x)}\,\theta^+\theta^-\cr
&\quad +
\left( ip\!\cdot\! \bar\d \bar x - i \bar p \!\cdot\! \bar\d x
- (p\!\cdot\! \bar\psi_R)(\bar p \!\cdot\! \psi_R)\right)
e^{i( p \cdot\bar x + \bar p\cdot x)}\,\bar\theta^+\bar\theta^-\cr
&\quad +
\left( ip\!\cdot\! \d \bar x - i \bar p \!\cdot\! \d x
- (p\!\cdot\! \bar\psi_L)(\bar p \!\cdot\! \psi_L)\right)\cr
&\qquad
\!\cdot\!\left( ip\!\cdot\! \bar\d \bar x - i \bar p \!\cdot\! \bar\d x
- (p\!\cdot\! \bar\psi_R)(\bar p \!\cdot\! \psi_R)\right)
e^{i( p\cdot\bar x + \bar p\cdot x)}\,
\theta^+\theta^-\bar\theta^+\bar\theta^- +\cdots \cr
&\equiv
O_p(z,\bar z)+J_p(z,\bar z)\theta^+\theta^-
+\bar J_p (z,\bar z)\bar\theta^+\bar\theta^-
+ V_p(z,\bar z)\theta^+\theta^-\bar\theta^+\bar\theta^-+\cdots\cr}
}
where the dots refer to terms with an odd  number of
$\theta$'s and/or $\bar\theta$'s (namely, terms whose holomorphic and/or
anti-holomorphic part is fermionic).  It will be useful in what follows
to factorize the various terms in eq.\eCOMP \ into left-- and
right--moving (holomorphic and anti--holomorphic) parts. Corresponding
respectively to the terms in \eCOMP , we denote the holomorphic and
anti-holomorphic vertex operators as follows:
\eqn\eCHI{
\eqalign{
V_p(X,\bar X)&\equiv
\O_p(z)\bar\O_p(\bar z)+ W_p(z)\bar\O_p(\bar z )\theta^+\theta^-
+\O_p(z)\bar W_p(\bar z)\bar\theta^+\bar\theta^-\cr
&\qquad+W_p(z)\bar W_p(\bar z)\theta^+\theta^-\bar\theta^+\bar\theta^-
+\cdots\cr}
}
where the exponential factors have been split using
$x(z,\bar z)=x_L(z)+x_R(\bar z)$. Thus, for example,
\eqn\eOV{
\eqalign{
\O_p(z)&=e^{i (p\cdot \bar x_L(z)+ \bar p\cdot x_L(z))}\cr
W_p(z)&=\left( ip\!\cdot\! \d \bar x - i \bar p
\!\cdot\! \d x - (p\!\cdot\! \bar\psi_L)(\bar p \!\cdot\! \psi_L)\right)
e^{i (p\cdot \bar x_L(z)+ \bar p \cdot x_L(z))}
\cr}
}
On the mass shell $p\cdot\bar p =0$ these are holomorphic vertex
operators of dimension 0 and 1, respectively, and the operators $O_p$,
$J_p$, $\bar J_p$, and $V_p$ appearing in eq.\eCOMP \ have dimensions
$(0,0)$, $(1,0)$, $(0,1)$, and $(1,1)$.  Each of these four types of
operators plays a distinct role in the vertex operator algebra.

The operators $O_p(z,\bar z)$ form a ring (the ``ground ring'') of
dimension $(0,0)$ operators.  They obey the multiplication law
\eqn\eOO{
O_p(z,\bar z)O_q(w, \bar w)= O_{p+q}(w, \bar w) + {\o}(|z-w|)
}
when $p$, $q$, and $p+q$ are all on shell.

The ground ring is acted upon by the algebra of currents $J_p(z,\bar
z)$ of dimension $(1,0)$ \it via \rm contour integration:
\eqn\eOJ{
\eqalign{\openup1pt
J_p(O_q)&\equiv {1\over 2\pi i}\oint_{C_w}
dz\, J_p(z,\bar z)\,O_q(w,\bar w)\cr
&\qquad= {1\over 2\pi i}\left(
\oint_{C_w} dz\, W_p(z)\O_q(w)\right) \bar \O_p(\bar
z)\bar \O_q(\bar w) +  \o(C_w) \cr
&\qquad
={1\over 2}(p\!\cdot\! \bar q-\bar p\!\cdot\! q) O_{p+q}(w,\bar w)
+ \o (C_w)\cr}
}
where the contour $C_{w}$ is taken to enclose $w$ once, and $\o(C_w)$
refers to elements that depend on the contour $C_w$.  A convenient
choice is to take $C_w$ to be an infinitesimal contour around $w$.
With that choice, $\o(C_w)$ can be shown to vanish for on--shell
momenta, and the currents $J$ act exactly as derivations on the ground
ring.

The $J$'s amongst themselves form a Lie algebra under the bracket
\eqn\eJJ{
[J_p,J_q]\equiv J_p(J_q)-J_q(J_p)=
(p\!\cdot\! \bar q - \bar p \!\cdot\! q)J_{p+q} + \o (C_w)
}
Completely analogous statements hold for $\bar J$ commutators as well.
The commutator of $J$ with $\bar J$ is in general non-zero and
will be discussed later.

As mentioned in the introduction, the $(1,1)$ operators $V_p(z,\bar
z)$ generate marginal deformations
\eqn\eMARG{
\eqalign{
S&=S_0+\delta S\cr
\delta S &=  \int d^2 z d^2\theta d^2 \bar\theta \int d^4p\,
\phi(p)\, V_p(Z,\bar Z)
=\int d^2 z  \int d^4p \,\phi(p)\,V_p(z,\bar z)\cr}
}
The operator product expansion (OPE) of the
marginal operators reproduces the 3-point function of eq.\eTHREE .
The currents $J$ act on the marginal operators as
\eqn\eJV{
J_p(V_q)=\half (p\!\cdot\! \bar q - q\!\cdot\! \bar p) V_{p+q} + \ldots
}
and likewise for $\bar J$ (the dots in \eJV \ refer to $(1,1)$
operators which are not upper components of $\O_{p+q}$).  A
transformation of $V_p$ is equivalent, by inspection of eq.\eMARG , to
a transformation of the couplings $\phi(p)$.  Thus, the
transformations \eJV \ relate different deformations of the action
$S_0$.

The algebras \eOJ , \eJJ , and \eJV \ are all dependent on the choice
of a contour for the nonholomorphic current $J_p$.  Because it is not
holomorphic, $J_p$ is not conserved, so it does not generate a true
symmetry.  However, as we will show in the next section, for any
on--shell $J_p$, there is always a particular choice of a compactified
background in which $J_p$ becomes holomorphic, and thus generates an
unbroken gauge symmetry.

Besides $O$, $J$, $\bar J$, and $V$, there are additional states at
discrete values of the momenta
\ref\rDIS{A.M. Polyakov, Mod. Phys. Lett.
{\bf A6} (1991) 635.}.  For the background $\R^{2,2}$, the only
discrete states are at $p=0$ \ref\rJB{J. Bie\'{n}kowska, preprint EFI
91-65, 1991.} \ (although we shall find many more when we consider
toroidal compactifications below). Vertex operators for dimension--one
discrete states are formed from off--shell $J_p$'s by choosing an
appropriate normalization as $p$ goes to zero:
\eqn\eDIS{
D_\eps(z)=i\eps\!\cdot\! \d \bar x - i \bar \eps \!\cdot\! \d x
=\lim_{\scriptstyle p\to 0 \atop \scriptstyle \hat p=\eps}
{1\over |p|} W_p(z)\bar\O_p(\bar z)
}
where $\eps_i$ is a constant complex two-vector of unit norm. The idea
here is to send $p$ to zero with a particular polarization, scaling
out the factor of $p$ appearing in $W_p$ that makes $W_p$ go to zero.
In order for the normalization $1/|p|$ to make sense, we must continue
to off--shell momenta (with $|p|\ne 0$) before taking the on--shell
limit $p\to 0$. Although the polarization $\eps_i$ must be non--null
in taking the limit in eq.\eDIS , arbitrary polarizations can  be
obtained by taking linear combinations of the $D_\eps$. We shall
refer to the vector components of $D_\eps$ and $\bar D_\eps$ as
\eqn\eDI{
D_i\equiv \d x^i\qquad\qquad \bar D_i\equiv \bar\d x^i
}
(where $i$ may be either a holomorphic or an antiholomorphic index).

The discrete operator $D_{\eps}$ is related by supersymmetry
transformations to the dimension zero operator $\O_\eps =
\eps\!\cdot\! \bar x - {\bar \eps} \!\cdot\! x$. However, $\O_\eps$
is not conformal, so
$D_{\eps}$ is not an upper component
of a lower--dimensional conformal field.
The holomorphic current $D_\eps$ (and likewise $\bar D_\eps$) generates
global isometries in the direction $\eps$.  As we shall see in the
next section, these discrete operators are naturally
incorporated into the operator algebra.

In addition to $D_i$, we can also form the discrete (1,1) operators
\eqn\eG{
V_{ij}(z,\bar z)=\d x^i \bar \d x^j
}
which generate deformations of the metric moduli.

It should be stressed that the operator algebras \eOO \ and \eJJ \
apply properly to {\it on--shell} momenta only, {\it i.e.}, $p\!\cdot\!
\bar p=q\!\cdot\! \bar q=(p+q)\!\cdot\! (\bar p + \bar
q)=0$.\footnote{$^*$}{If $(p+q)\!\cdot\! (\bar p + \bar q)< 0$, the
$\o(C_w)$ in \eOJ, \eJJ \ can not be set to $0$ by a choice of a contour.
Declaring that $\o(C_w)=0$ anyway will lead to an algebra that
violates the Jacobi identity.} Strictly speaking, then, in defining
these algebras $O_p$, $J_p$, and $V_p$ should be indexed by momenta
restricted to a subspace $V_{\rm null}$ of null vectors, such that the
sum of any two momenta $p,q\in V_{\rm null}$ is also null: $p\!\cdot\!
\bar q + \bar p \!\cdot\! q=0$. A maximal such subspace of $\R^{2,2}$ is
spanned by any two perpendicular null vectors.  We will refer to
algebras of operators with momenta lying in a null plane of this sort,
as on--shell algebras \ref\rOSA{V.A. Kosteleck\'y and O. Lechtenfeld,
Phys. Rev. Lett.\ \bf 59 \rm (1987) 169\semi R.E. Borcherds, Proc.
Natl. Acad. Sci.USA \bf 83 \rm (1986) 3068\semi S. Nergiz and C.
Saclioglu, Int. J. Mod. Phys.\ \bf A5 \rm (1990) 2647.}.

A nondegenerate triple of on--shell momenta $(p, q, -(p+q))$
determines a unique null plane.  The spacelike and timelike components
of $(p,q,-(p+q))$ may each be thought of as forming the edges of two
congruent triangles \rOV, which may have equal or opposite
orientation. If the triangles' orientations are equal, then the
corresponding 3--point function vanishes, while for opposite
orientations it is nonvanishing. Thus to specify a pair of momenta in
this two--dimensional null space it is sufficient to specify only
their (complex) spacelike components $p_s$, $q_s$, if we are
interested in the nontrivial part of the operator algebra. Let us
restrict attention to this nontrivial subalgebra and accordingly let
us label its generators by the unrestricted spacelike momenta $p_s$.
Then the on--shell algebra \eJJ\ can be written in terms of the $p_s$
as
\eqn\eON{
[J_{p_s},J_{q_s}]=2(p_s{\bar q}_s-q_s\bar p_s) J_{p_s+q_s}
}
Up to a rescaling, this is nothing but the Lie
bracket in the algebra of area--preserving diffeomorphisms of the plane
as generated by the basis
\eqn\eLDEF{
L_{k}\equiv ie^{i(k\bar x + \bar k x)}( \bar k \d_{\bar x} - k \d_x )
}
In fact,
\eqn\eAPD{
[L_k,L_{k'}]= (k{\bar k}'-k'\bar k)L_{k+k'}
}
so the identification of $J_k$ with $2L_k$ gives a Lie algebra
isomorphism between \eAPD \ and \eON .  Geometrically, the $J$'s
generate area--preserving diffeomorphisms of the null plane $V_{\rm
null}$, and act on the ring of normalizable functions on $V_{\rm
null}$, which is the ground ring generated by the $\O_p(z,\bar z)$.

We conclude that area--preserving diffeomorphisms of the null plane
are a subalgebra of the off--shell algebra of vertex operators.
Indeed, area--preserving diffeomorphisms of \it any \rm null plane are
contained in the full algebra. As we will argue in the next section, a
natural candidate for this off--shell algebra is the algebra of
volume--preserving diffeomorphisms of the target space, vdiff$(M)$.

\newsec{Toroidal backgrounds}

So far, we have just considered the background $\R^{2,2}$. A
straightforward generalization is to compactify some or all of the
coordinates --- including timelike dimensions --- on a torus.
Compactification of timelike dimensions is quite unphysical, but will
prove to be a useful trick in determining an underlying gauge algebra.
Namely, for each on--shell subalgebra, we will be able to find a
background in which that symmetry is unbroken.  More precisely, what
we will find is that for every null plane algebra of the form \eJJ ,
there is a point in the moduli space of toroidal compactifications at
which the currents become holomorphic. At this point, the currents
$J_p$ and $J_q$ are conserved, their definitions as contour integrals
\eOJ \ become independent of the contours chosen, and the gauge
symmetry is realized exactly on physical states.

The most general toroidal compactification is constrained by the
level--matching requirement, which must be satisfied by both on-- and
off--shell states:
\eqn\eTOR{
\left( |p_{Ls}|^2 - |p_{Lt}|^2\right) -\left(
|p_{Rs}|^2-|p_{Rt}|^2 \right) \in 2\Z
}
This condition is satisfied by the vectors of an even Lorentzian
lattice of signature $(4,4)$.\footnote{$^\dagger$} {The signature
$(4,4)$ should be regarded as a notational shorthand for $(2,2;2,2)$,
where the semicolon separates the signatures of the left and right
movers.}\ Modular invariance of the
formal 1-loop partition function also
constrains the lattice to be self--dual, so we just have a Narain
compactification \rNAR \ on $\Gamma^{4,4}$.

We begin by deriving the on--shell algebra. The on-shell condition
requires that the two expressions in parentheses in eq.\eTOR \ be
separately zero --- namely, that the on--shell vectors generate a
subset of the null vectors of $\Gamma^{4,4}$.  Sublattices of
$\Gamma^{4,4}$ of mutually perpendicular null vectors are at most
four--dimensional.  It is thus advantageous to study
compactifications for which  all
the vectors in such a null sublattice are on--shell,
in order to describe a maximal
on--shell algebra.  In fact, there always exists
a four--dimensional subspace of
mutually perpendicular on--shell null momenta if the
(4,4)--dimensional lattice is a direct sum of two (2,2)--dimensional
ones
\eqn\eSUM{
\Gamma^{4,4}= \Gamma^{2,2}_L \oplus \Gamma^{2,2}_R
}
On--shell momenta $p_L$ and $p_R$ may then be taken to lie in
two 2--dimensional null sublattices  and vertex operators
carry indicies $(p_L, p_R)$. If we wanted to, we could as before label
on--shell momenta (giving rise to non-trivial 3-point functions)
by their spacelike components only; however, in
order to allow generalization to off--shell momenta, we will take
$(p_L,p_R)$ to represent the full $(4,4)$-dimensional lattice vector.

Recall that we are interested in any $(1,0)$ (or $(0,1)$) operator
that may become holomorphic (anti-holomorphic) somewhere in the moduli
space of toroidal backgrounds.  All operators with momenta which are
null with respect to the Lorentzian norm in eq.\eTOR \ have this
property. To see this, let $p=(p_L,p_R)$ be any null lattice vector in
$\Gamma^{4,4}$.  By rotating $\Gamma^{4,4}$ by the set of $SO(4,4)$
transformations we can cover the full moduli space of (4,4) lattices.
Now $SO(4,4)$ transformations act on individual 8--real--component
momenta $(p_{Ls},p_{Rt},p_{Lt},p_{Rs})$ in the fundamental
representation, and an $SO(4,4)$ transformation can always be found,
which rotates $p$ into a null vector of the form
$(p_{Ls}',0,p_{Lt}',0)$.  That is, there always exist points in moduli
space where $J_p=J_{(p_L,0)}$ is holomorphic. At such a point, $J$ is
exactly conserved, and the charge given by the contour integral of $J$
will be independent of the choice of contour.  Indeed, more is true:
given two operators $J_p$ and $J_q$ with $p$, $q$, and $p+q$ all
on--shell, there is always an $SO(4,4)$ rotation taking $p$ and $q$
{\it simultaneously} to $(p_L,0)$ and $(q_L,0)$. This is seen by first
rotating $p$ by a particular $SO(4,4)$ transformation. The
subgroup of $SO(4,4)$ preserving $p$ may then be used to rotate the
orthogonal vector $q$ into the desired form. Hence there is always a
point in the moduli space of toroidal backgrounds where the Lie
bracket \eJJ \ is exact, independent of the choices of contours for
$J_p$ and $J_q$.  Once we move away from the special point where the
currents are holomorphic, the corresponding gauge symmetries will be
spontaneously broken.  Taken together, the on--shell currents generate
an enormous symmetry algebra of the full theory, which is broken down
to a subalgebra by any given toroidal background.

Even the unbroken subalgebra in a particular background may be
infinite--dimensional. For example, in a toroidal background of the
type \eSUM \ there are an infinite number of holomorphic (and
anti-holomorphic) currents, corresponding to momenta of the form
$(p_L,0)$ (and $(0,p_R)$). These currents give rise to {\it exact}
infinite symmetry algebras of the CFT: the area--preserving
diffeomorphisms of null 2-tori in the $(2,2)_L$ (and $(2,2)_R$)
torus.

For a general toroidal compactification,
the on--shell algebra generated by the $J$ and $\bar J$ is
\eqn\eSYM{
\eqalign{
[J_{(p_L,p_R)},J_{(q_L,q_R)}]&=
(p_L\!\cdot\!{\bar q}_L-q_L\!\cdot\!{\bar p}_L) J_{(p_L+q_L,p_R+q_R)}\cr
[{\bar J}_{(p_L,p_R)},{\bar J}_{(q_L,q_R)}]&=
(p_R\!\cdot\!{\bar q}_R-q_R\!\cdot\!{\bar p}_R)
{\bar J}_{(p_L+q_L,p_R+q_R)}\cr
[J_{(p_L,p_R)},{\bar J}_{(q_L,q_R)}]&=
\half(p_L\!\cdot\!{\bar q}_L-q_L\!\cdot\!{\bar p}_L) {\bar
 J}_{(p_L+q_L,p_R+q_R)}\cr
&\qquad+ \half(p_R\!\cdot\!{\bar q}_R-q_R\!\cdot\!{\bar p}_R)
J_{(p_L+q_L,p_R+q_R)}+ ...\cr}
}
where the dots refer to $(1,0)$ and $(0,1)$ operators which are not
upper components of $\O_{(p_L+q_L,p_R+q_R)}$.  The OPE of the $V$'s is
\eqn\eQA{
\eqalign{
&V_{(p_L,p_R)}(z,\bar z)\cdot V_{(q_L,q_R)}(w,\bar w)=\cr
\openup5pt
&\qquad\qquad
\hbox{$1\over 4$}{(p_L\!\cdot\!{\bar q}_L-q_L\!\cdot\!{\bar p}_L)
(p_R\!\cdot\!{\bar q}_R-q_R\!\cdot\!{\bar p}_R)\over |z-w|^2}
\,V_{(p_L+q_L,p_R+q_R)}(w,\bar w)+\cdots\cr}
}

By a simple extension of eq.\eLDEF, we can obtain
a representation of the algebra of the $J$'s and $\bar J$'s as
follows. Let
\eqn\eLLR{
\eqalign{
L_{(p_L,p_R)}^{\pm}&=J_{(p_L,p_R)}\pm {\bar J}_{(p_L,p_R)}\cr
\openup5pt
&\equiv
ie^{i(p_L {\bar x}_L+{\bar p}_L x_L+p_R {\bar x}_R+{\bar p}_R x_R)}
[({\bar p}_R \!\cdot\!\d_{{\bar x}_R} - p_R \!\cdot\!\d_{x_R})
\pm ({\bar p}_L \!\cdot\!\d_{{\bar x}_L} - p_L\!\cdot\!\d_{x_L})]\cr}
}
The $L^+$ and $L^-$ act on the ground ring of functions on the Narain
$(4,4)$--torus as derivations, \it i.e.\rm, diffeomorphisms.  In fact,
the $L^+$ and $L^-$ generate algebras of {\it symplectic} diffeomorphisms.
Before describing these algebras, we should briefly recall some basic
facts about symplectic geometry.

Let $\omega$ be a closed 2--form on a $2n$--dimensional
manifold $M$ whose $n$th power is proportional to the volume form
of $M$, and let $f$
denote any differentiable function on $M$. Then there is a vector
field $v_f$ associated with $f$,  whose interior product  with
the symplectic form $\omega$ is $df$:
\eqn\eSD{
i(v_f)\,\omega= df
}
Such a vector field is said to be symplectic with respect to $\omega$,
and the Lie derivative of $\omega$ in the direction $v_f$ is
automatically 0; that is, the flow generated by $v_f$ preserves
$\omega$.  The symplectic vector fields form the Lie algebra of
symplectic diffeomorphisms sdiff$_\omega$(M), whose Lie bracket may be
shown to satisfy
\eqn\eLB{
[v_f,v_g]=v_{\{ f,g\} }
}
where $\{ f, g \} = \omega(v_f,v_g)$
is the Poisson bracket with respect to $\omega$.
Since the $n$th power of $\omega$ is proportional to the
volume form on $M$, a symplectic diffeomorphism is automatically
volume--preserving.
(There may be additional diffeomorphisms preserving $\omega$; these
correspond to closed 1-forms on $M$ which are not derived from any
$f$. Such 1-forms are precisely the elements of $H^1(M)$.)

Any K\"ahler manifold comes equipped with a natural symplectic
structure, given simply
by the K\"ahler form. The Narain $(4,4)$ torus is a K\"ahler
manifold with K\"ahler form (in an appropriate basis)
\eqn\eKF{
k=dx_L^1\w d\bar x_L^1
-dx_L^2\w d\bar x_L^2
-dx_R^1\w d\bar x_R^1
+dx_R^2\w d\bar x_R^2
}
corresponding to the Lorentzian metric on $T^{4,4}$ implicit in \eTOR .
It is with respect to this $k$ that the $L^+$ are symplectic. Indeed,
the symplectic vector field
\eqn\eLK{
L^+_f \equiv k^{i j} \d_i f \d_j
}
(where $k^{ij}$ is the inverse of the matrix $k_{ij}$ representing
$k$) readily reduces to $L^+_p$ when $e^{ip\cdot x}$ is substituted
for $f$.  Thus, for on--shell momenta the $L^+_p$ generate a null
subalgebra of sdiff$_k(T^{4,4})$.

The $L^-$ also close on themselves as an algebra of symplectic
diffeomorphisms relative to a different symplectic form
\eqn\eKF{
\tilde k=dx_L^1\w d\bar x_L^1
-dx_L^2\w d\bar x_L^2
+dx_R^1\w d\bar x_R^1
-dx_R^2\w d\bar x_R^2
}

In addition,
the Lie bracket of an $L^+$ with an $L^-$ will generate
volume--preserving diffeomorphisms symplectic with respect to still
other forms:
\eqn\eLL{
[L^+_f,L^-_g]
=(k^{ij}\tilde k^{kl}
-k^{il}\tilde k^{kj})
\d_j(\d_i f\d_k g)\d_l
}
The right side can always be decomposed in terms of symplectic forms
on $T^{4,4}$
\eqn\eDECOMP{
k^{ij}\tilde k^{kl} -
k^{il}\tilde k^{kj}
=\sum_I A_I^{ik} \omega^{jl}_I
}
where $I$ indexes a basis of symplectic matrices on $T^{4,4}$.  (This
decomposition follows from the fact that any antisymmetric matrix can
be written as a sum of symplectic matrices.)  In terms of the
$\omega_I$, we may rewrite eq.\eLL \ as
\eqn\eLLO{
[L^+_f,L^-_g]=\sum_I \omega_I^{jl}\,\d_j (A_I^{ik}\d_i f \d_k g) \d_l
}
which makes it clear that the Lie bracket is always a sum of
symplectic diffeomorphisms.   The additional symplectic
diffeomorphisms thus generated correspond to the dots in eq.\eSYM .

Besides the above operators, we can again form discrete states
$D_{\eps_L}(z)$ and ${\bar D}_{\eps_R}(\bar z)$ as in \eDIS,
corresponding to the $U(1)_L^4\times U(1)_R^4$ isometries of the
toroidal background. These states combine naturally with the
symplectic diffeomorphisms. In fact, they are precisely the extra
volume--preserving diffeomorphisms which are not symplectic with
respect to any $\omega$, and are generated by vector fields $v$ for
which the left side of eq.\eSD \ is closed but not exact.

The geometric interpretation of the zero--momentum discrete states
merits a brief digression.  The dimension $(1,1)$ operators $V_{ij}$
appearing in eq.\eG \ generate modular deformations of
(4,4)--dimensional Narain compactifications \ref\rNSW{K.S. Narain,
M.H. Sarmadi and E.  Witten, Nucl. Phys. {\bf B289} (1987) 414.}.  As
such they are naturally associated with elements of $H^{(1,1)}(M)$.
In a more general curved background, the zero--momentum condition on
the dimension $(1,1)$ discrete operators becomes the condition that
the $(1,1)$--form corresponding to the associated compactification
modulus be harmonic.  Thus there is a correspondence between discrete
operators of worldsheet dimension $(1,1)$ and closed $(1,1)$ forms on
$M$.  Likewise, the dimension--one operators $D_{\eps_L}$ and ${\bar
D}_{\eps_R}$ generating isometries of the Lorentzian torus are also
related to the cohomology of $M$. In general, the discrete states of
dimension one will be associated to nonsymplectic volume--preserving
diffeomorphisms of $M$, which are in one-to-one correspondence with
the elements of $H^1(M,\R)$.
\rm

Actually, now that we are working in a compact
background there are many more discrete (1,0) and (0,1) states, at
momenta $(0,p_R)$ and $(p_L,0)$ with $p_L^2=p_R^2=0$ (here $(p_L,p_R)$
includes the time-like components). The corresponding (1,0) operators
$D_{(\eps_L, p_R)}(z,\bar z)\equiv D_{\eps_L}(z)\bar \O_{p_R}(\bar z)$
act on the $J$'s as
follows:
\eqn\eDJ{
[D_{(\eps_L,p_R)}, J_{(q_L,q_R)}]
=(\eps_L \!\cdot\! {\bar q_L} -{\bar \eps}_L \!\cdot\! q_L) J_{(q_L,p_R+q_R)}
}
The complete underlying algebra generated by $J, {\bar J}, D$ and ${\bar
D}$ is given by their representations as derivations:
\eqn\eDD{
D_{(\eps_L,p_R)}\equiv ie^{i(p_R\!\cdot\!
{\bar x}_R+{\bar p}_R \!\cdot\! x_R)}
({\bar \eps}_L \d_{{\bar x}_L}-\eps_L\d_{x_L})
}
and similarly for ${\bar D}_{(p_L,\eps_R)}$.  We immediately recognize
$D_{(\eps_L,0)}=D_{\eps_L}$ and ${\bar D}_{(0,\eps_R)}=\bar
D_{\eps_R}$ as  generators of isometries of the torus.

The other $D_{(\eps_L,p_R)}$,
with $p_R\neq 0$, are actually symplectic diffeomorphisms, with
respect to symplectic forms that mix left and right movers, of the form
\eqn\eDLR{
\omega_{ij}\,dx_L^i\w dx_R^j
}
If coordinates are chosen so that the vector $\eps_L$ is dual to the
one-form $dx^1_L$ and $p_R$ is dual to $dx^1_R$, then a symplectic
structure generating $D_{(\eps_L,p_R)}$ is
\eqn\eWLR{
\omega_{(\eps_L,p_R)}
=dx_L^1\w dx_R^1 + (\hbox{terms $\perp$ to $\eps_L$ and $p_R$})
}
Explicitly, we obtain
\eqn\eDF{
D_{(\eps_L,p_R)}=\omega^{ij}_{(\eps_L,p_R)}\d_i f(x_R)\d_j
}
with the  function $f(x_R)=e^{ip_R\cdot x_R}$.
The discrete operators $D_{(\eps_L,p_R)}$ are already included in the
algebra generated by the $L^+$ and $L^-$, as is seen by taking, for
example, $f=f(x_R)$ and $g=g(x_R)$ to be independent of $x_L$ in
eq.\eLL .

\bigskip

The representation of the on--shell algebra as a
derivation algebra turns out to be more than a convenient encoding; it
also suggests a natural  underlying off--shell  algebraic extension.

The generalization of the algebras in \eSYM \ and \eDJ \ to off--shell
momenta begins with the observation that there is an on--shell algebra
associated with any null sublattice $\Gamma_{\rm null}$.  There are
many possible choices of $\Gamma_{\rm null}$, and the full algebra
should contain all the resulting on--shell algebras.  A natural
candidate for this off--shell algebra is the derivation algebra
generated from \eLLR \ and \eDD \ over the set of {\it all} null
momenta of the (4,4) lattice.  Since in fact there is a basis for
$\Gamma^{4,4}$ consisting entirely of null momenta, we immediately
obtain generators for each lattice momentum $(p_L,p_R)$ as Lie
brackets of on--shell generators $L^\pm$. The $L^+$ and $L^-$ each
generate algebras of symplectic diffeomorphisms sdiff$_k(T^{4,4})$ and
sdiff$_{\tilde k}(T^{4,4})$. Together with the $D$'s they generate the
full volume--preserving diffeomorphism algebra vdiff$(T^{4,4})$.
(This is a nontrivial statement since there are 105 independent
symplectic structures on $T^{4,4}$; it can be checked by taking linear
combinations of the derivations appearing on the right--hand side of
\eLL .)  The off--shell algebraic extension to vdiff$(T^{4,4})$ is
background--independent, because all the Narain tori are isomorphic.

In the standard decompactification limit, where $p_L=p_R$, the full
algebra reduces to the algebra of the $L^+$'s. The $L^-$ act trivially
on themselves and on the ground ring of functions on the target space
$\R^{2,2}$, and the $L^+$ generate sdiff$(\R^{2,2})$.

It is also interesting to consider the case where only
the spacelike dimensions are compactified. Now the on-shell condition
requires that $(p_{Ls},p_{Rs})$ be a null vector in $\Gamma_s^{2,2}$:
$(p_{Ls})^2-(p_{Rs})^2=0$ (because $p_{Lt}=p_{Rt}$ for
non-compact time). Since a maximal sublattice $\Gamma_{\rm null}$ of
the Narain lattice $\Gamma_s^{2,2}$ is two--dimensional, the on--shell
algebra in this case reduces to the area-preserving diffeomorphism
algebra of a two-dimensional null torus. The underlying off--shell
algebra generated by the derivations is isomorphic to
vdiff($T_s^{2,2}$), namely, to the volume--preserving diffeomorphisms
of the Narain torus of this compactification.

\bigskip

We conclude this section by pointing out some similarities between the
algebraic structures of the $N=2$ and $d=2$ strings.

The on--shell algebra of the $N=2$ string contains the on--shell
algebra of the $d=2$ string in a trivial way.  To see this, let
$p^{(1)}_{L}$ and $p^{(2)}_{L}$ be two basis vectors for the
left--moving two--dimensional null sublattice of $\Gamma_L$ in \eSUM,
and similarly choose a basis for the right--movers. Following
\ref\rGR{E. Witten, preprint IASSNS-HEP-91-53, 1991.},
define the holomorphic (and anti-holomorphic) dimension 0 operators
\eqn\eXY{
\eqalign{
x&=\O_{p^{(1)}_L}(z)\cr
y&=\O_{p^{(2)}_L}(z)\cr
x'&=\bar\O_{p^{(1)}_R}(\bar z)\cr
y'&=\bar\O_{p^{(2)}_R}(\bar z)\cr}
}
With this choice, a subring of the ground ring is generated by
polynomials in the four variables
\eqn\eAI{
a_1=xx',\qquad a_2=yy',\qquad a_3=xy', \qquad a_4=yx'
}
and their inverses, with the relation
\eqn\eCONE{
a_1a_2-a_3a_4=0.
}
The algebra \eSYM \ acts on the 3-dimensional cone of such $a_i$.
This is quite similar to the situation studied in ref.\rGR , where the
symmetry algebra of the $d=2$ string (at the self--dual point) acts as
the algebra of volume--preserving diffeomorphisms of the cone \eCONE.

In case $p_L$ and $p_R$ both live in the same
Narain lattice $\Gamma^{2,2}$ (namely,
the symmetric toroidal compactification $\Gamma_L^{2,2}
=\Gamma_R^{2,2}$) the choice of
basis \eXY \ is more restricted. At the self-dual point (with
$SU(2)^2_L \times SU(2)^2_R$
extended symmetry), a good choice is
\eqn\ePPPP{
p_{Ls}^{(1)}=p_{Rs}^{(1)}={1\over \sqrt{2}}(1,1)\,;\qquad
p_{Ls}^{(2)}=p_{Rs}^{(2)}={1\over \sqrt{2}}(1,-1)\, .
}
With this choice,
the OPE \eQA \ coincides with the algebra found
in \ref\rKP{I. Klebanov and A.M. Polyakov,
Mod. Phys. Lett. {\bf A6} (1991) 3273.} for the $d=2$ closed string at
the self-dual point.
This can be shown as follows: amongst the vectors
$(p_{Ls}^{(i)};p_{Rs}^{(j)})$ only three (denoted $v^i$) are independent
\eqn\eVVV{
v^1={1\over \sqrt{2}}(1,1;1,1)\,;\qquad
v^2={1\over \sqrt{2}}(1,-1;1,-1)\,;\qquad
v^3={1\over \sqrt{2}}(1,1;1,-1)\,.\qquad
}
On--shell momenta are now labeled  by the space-like components
\eqn\ePQ{
p_s=\sum_{i=1}^3 p_i v^i\,;\qquad q_s=\sum_{i=1}^3 q_i v^i
}
where $p_i,q_i$ are integers.
Denoting $V_p= V_{(p_1,p_2,p_3)}\equiv V_{j_p,m_p,m'_p}$ with
$j_p=\half(p_1+p_2+p_3),\,m_p=\half(p_1-p_2+p_3),\,m'_p
=\half(p_1-p_2-p_3)$, one finds for the OPE \eQA :
\eqn\eVJVJ{
\eqalign{
\openup3pt
&V_{j_p,m_p,m'_p}(z,\bar z) \cdot V_{j_q,m_q,m'_q}(w,\bar w)\cr
&\qquad={(j_pm_q-j_qm_p)(j_pm'_q-j_qm'_p)\over |z-w|^2}
V_{j_p+j_q,m_p+m_q,m'_p+m'_q}(w,\bar w)+\cdots}
}
reproducing the algebra of \rKP .

The remarkable similarity between the $d=2$ and $N=2$ string algebras
is suggestive of a deep connection between the two theories, which
should be investigated further.

The full off--shell operator algebra is of course harder to obtain
directly.  We have already discussed a good candidate; yet it is far
from unique. In fact, another straightforward off--shell extension of
the on--shell algebras is a Lorentzian lattice algebra, of the type
discussed in \rGP . Further arguments will be required to justify our
choice of an off--shell algebra. The justification will come in the
following section when we consider gauge symmetries from the point of
view of the spacetime action.

\newsec{Spacetime action}

In the previous section, we constructed on--shell generators of exact
gauge algebras in different backgrounds, and found a natural
off--shell Lie algebraic closure in the algebra of volume--preserving
diffeomorphisms of the target Narain torus.  Here we present
independent evidence that vdiff$(T^{4,4})$ is an underlying gauge
symmetry algebra of the $N=2$ string.

A spacetime action leading to the correct on--shell
amplitudes in uncompactified Minkowski space is \rOV\
\eqn\eACT{
S_\phi=\int d^2 x_1 d^2 x_2 \left[
\half\eta^{i\bar j}\d_i \phi \d_{\bar j} \phi
+ \hbox{$1\over 3$}  \eps^{ij}\eps^{\bar i \bar j}
\phi \d_j \d_{\bar i}  \phi
\d_i \d_{\bar j} \phi\right].
}
This action implies the three--point  function
\eqn\eOFF{
\langle V_{p}V_{q}V_{r}\rangle =
-4(q\!\cdot\! \bar r)(\bar q \!\cdot\! r)+ 4 (q\!\cdot\! \bar q )
(r \!\cdot\! \bar r)
}
which reduces to eq.\eTHREE\ when all momenta are on--shell.

We can generalize the action \eACT \ to an arbitrary background, in a
way that makes the off--shell symmetry manifest, as follows.  From the
K\"ahler potential and the metric we can construct
\eqn\eL{
S=\third\int K\w k \w k = \third\int d^4 x \sqrt g K
}
where $k$ is the K\"ahler (1,1)-form derived from the (0,0)-form $K$
and $g$ is to be treated as a function of $K$.  This is essentially
the unique action (up to total derivatives and,
as we shall explain, terms arising from the existence of discrete
states), which involves only the K\"ahler potential and its
derivatives, is invariant under K\"ahler transformations, and
reproduces all 3- and higher--point correlation functions when
expanded around flat Minkowski space.  Upon substitution of $K_0+\phi$
for $K$ (where for example $K_0$ may be taken to be the K\"ahler
potential for $\R^{2,2}$; $K_0=x_1{x}_{\bar 1}-x_2{x}_{\bar 2}$) this
action reverts to the cubic term in Ooguri--Vafa action
\eACT, plus a (somewhat arbitrary) quadratic term, a tadpole
and a $\phi$--independent term:
\eqn\eTAD{
S=S_\phi+\int  \left(\half \phi\d\bar \d \phi \w\d\bar\d K_0
+ \phi\d\bar \d K_0 \w\d\bar\d K_0
+\third K_0\d\bar \d K_0 \w\d\bar\d K_0\right)
}
Of course, if $K_0$ is a solution of the equations of motion, the
tadpole will vanish.  Varying within the class of K\"ahler metrics --
{\it i.e.}, with respect to $K$ -- we obtain the equation of motion
\eqn\eEOM{
\det g = 0
}
The solutions to \eEOM \ lie on the boundary of the moduli space of
K\"ahler geometries, and the functional measure for the action \eL\ is
therefore peaked about singular metrics, just as in topological
theories of gravity. We will shortly see how to modify the action in
order to obtain nonsingular backgrounds as solutions to the equations
of motion as well.

The symmetries of the action \eL \ are manifest; it
is invariant under diffeomorphisms that preserve the K\"ahler form
$k$. Such diffeomorphisms clearly leave the volume element $k\w k$
invariant, and a diffeomorphism that preserves $k$ can only change $K$
by a K\"ahler gauge transformation. Under a K\"ahler transformation
$K(x,\bar x) \to K(x,\bar x)+{\rm Re}\,f(x)$, the Lagrangian changes
by a total derivative, as seen by integrating by parts twice.
(The same sort of gauge invariance up to a total derivative is also
a property of  the Chern--Simons action in $2\!+\!1$ dimensions.)
Assuming that the total derivative can be ignored, this shows that
$k$--symplectic diffeomorphisms are indeed symmetries.

These are precisely the
symmetries generated by the $L^+$ in eq.\eLLR, which formed the
surviving part of the off--shell algebra in the limit of
decompactification to $\R^{2,2}$.

\bigskip

Because of the presence of a tadpole, the action \eL \ is not
equivalent to the action \eACT \ derived by Ooguri and Vafa.  However,
there is a term we can add to it, which cancels the tadpole without
violating any symmetries. This is the quadratic term
\eqn\eKT{
\half\int d^4 x\, G^{i\bar j} K\d_i \d_{\bar j} K
=-\half\int d^4 x\, G_{j\bar i} \epsilon^{ij}\epsilon^{\bar i \bar j}
K \d_i \d_{\bar j} K
}
For constant $G_{i\bar j}$, this term is invariant under symplectic
diffeomorphisms preserving $k$.
Adding it to the background--independent action $S$, we obtain
\eqn\eSTOT{
S_G=\int  \left(\third K\w k \w k -\half  K\w k \w k_G \right) }
where $K_G$ is the K\"ahler potential for $G$ and $k_G$ is the
(1,1)--form derived from $K_G$.

With this modified action, the equation of motion for $K$ is
\eqn\eNEWEOM{
\dd K \w \dd K = \dd K \w \dd K_G
}
which is solved (for example) by $K=K_G$.  In other words, $K_G$
should be regarded as the classical background $K_0$, in order for the
tadpole to vanish.  Expanding around $K_0=K_G= \eta_{i\bar j}x^i
x^{\bar j}$, we recover exactly the original action \eACT \ (plus a
$\phi$-independent term).\footnote{$^*$}{Another solution
to the equation of motion \eNEWEOM \
is $K=0$. Expanding $K$ around this classical solution, we also
recover the action \eACT \ if $K_G=-\eta_{i\bar j}x^i x^{\bar j}$.
(Note that the metric $-\eta_{i\bar j}$ is equivalent to $\eta_{i\bar
j}$ in signature (2,2).)}\ \ The equation of motion then becomes
\eqn\ePHIEOM{
\dd \phi \w \dd \phi + \dd K_0 \w \dd\phi = 0
}
as found in \rOV. This is equivalent to the Plebanski equation
\ref\rPLEB{C. Boyer and J. Plebanski, J. Math. Phys.\
\bf 26 \rm (1985) 229.}.

There is another reason for adding the term \eKT , besides cancelling
the tadpole, involving the existence of discrete states.  In a flat
Minkowski background, for example, there are discrete states
corresponding to the 0--momentum graviton vertex operator
\eqn\eGVO{
V^{i\bar j} =\d x^i \bar \d  x^{\bar j} + \bar \d x^i \d x^{\bar j}
}
We associate this operator with a constant background metric $G_{i\bar
j}$ derived from the K\"ahler potential $\tilde K_0= G_{i\bar j}x^i
x^{\bar j}$. For convenience, we may choose coordinates so that
$\tilde K_0 = x^1  x^{\bar 1} - x^2 x^{\bar 2}$.

Now there is a nonvanishing 3--point function involving the
graviton discrete state \eGVO
\eqn\eGPP{
\langle V_{i\bar j} V_p V_{-p}\rangle=
p_i {p}_{\bar j}+{p}_{\bar i} p_j
}
The tadpole--cancelling term \eKT \ can therefore be identified as a
contribution of the discrete states to the background--independent
effective action in \eL .

The discrete states $D^i$ of conformal dimension one also play a
role in the spacetime action.
Consider making the following K\"ahler transformation
\eqn\eLKT{
K(x,\bar x)\to K(x,\bar x)+ a_i x^i + b_{\bar i} x^{\bar i}
}
where the $a_i$ and $b_i$ are complex constants.  Such a
transformation leaves the metric and the equations of motion
unchanged, but shifts the one--form $dK$ by $a_i dx^{i}+ b_{\bar i}
dx^{\bar i}$.  Recall from section 2 that the discrete state $D^i$ is
an upper component of the non--normalizable field $x^i$; the shift of
$K$ in \eLKT \ is therefore associated with $D^i$ and $D^{\bar i}$.
The effect of the K\"ahler transformation \eLKT \ on the action \eSTOT
\ is to introduce a background gauge potential $A=a_i dx^i +b_{\bar i}
dx^{\bar i}$
\eqn\eSWIND{
S_{G,A}= \int \left[ -\third
(\d K+A)\w (\bar \d K+ \bar A) \w \dd K
+\half (\d K+A)\w (\bar \d K+\bar A) \w k_G \right]
}

The action
\eSTOT \ (derived for a flat Minkowski background)
may be extended to more general
backgrounds.
The simplest extension is to complexified K\"ahler structures
\ref\rCK{S.\ Ferrara and A.\ Strominger, in Strings '89 Workshop
proceedings, College Station, TX.}, with nonvanishing constant
two--form field $B_{i\bar j}$. In such backgrounds, it is natural to
work with a complex K\"ahler potential, satisfying
$\d_i\d_{\bar j} K = G_{i\bar j}+ iB_{i\bar j}$.  The action \eSTOT \
then becomes complex. The on--shell $3$--point amplitudes are real in
Minkowski space, and are unaffected if we take the real part of this
complex action.

The generalization to compact backgrounds is more problematic.
We derived the action \eSTOT \ from on--shell amplitudes in a
noncompact Minkowski background, and we might not expect it to apply
in a toroidal background.  On a torus, the off--shell three--point
function \eOFF\ no longer admits an obvious decomposition into
left-handed and right-handed parts; it is hard to see how a
generalization of the intrisically 4--dimensional action \eL \ could
produce the compactified on--shell 3-point amplitude contained in
\eQA .  After all, the Narain torus is 8--dimensional, and momenta have
separate left-- and right--moving components, which are not reflected
in the momentum spectrum of $K$.

Alternatively, we may choose to work in the 4--dimensional picture of
Narain, Sarmadi, and Witten \rNSW , but then we must figure out how to
include winding modes. From the point of view of a spacetime action,
these would be nonlocal excitations. We might hope to find them
appearing as solitons, as cosmic strings wrapped around the spacetime
torus \ref\rCSS{E.\ Witten, Phys. Lett.\ \bf B153 \rm (1985) 243.}.
If the winding modes of the string are actually solitons of the field
$K$, then it may be that the spacetime action \eSWIND \ is already
complete for any toroidal background. Shifting the winding sector $w$
of $K$ would allow us to obtain a state of any $p_L=p+w$ and
$p_R=p-w$. Then the additional symmetries not contained in the
action's manifest symmetry algebra sdiff$_k (M)$ would act by mixing up
local momentum excitations of $K$ with solitonic winding states.  An
analogous situation arises in $N=2$ super Yang--Mills theories
\ref\rWO{E.  Witten and D.  Olive, Phys. Lett.\ {\bf B78} (1978) 97.},
where the full theory may have an extra duality symmetry relating
local electric and nonlocal solitonic magnetic monopoles, which
however is not manifest in the local action.
\bigskip

To recapitulate, we have found an action that gives back the correct
on--shell amplitudes and is invariant under the algebra of symplectic
diffeomorphisms, which coincides with the derivation algebra derived in
section 3 from the worldsheet point of view.  The discrete states
played a key role in the construction: they contributed a
background--dependent term that allowed us to shift the tadpole away
in a flat Minkowski background.

The effective action was derived in two stages. First, following
ref.\rOV, we wrote down the minimal Lorentz--invariant action
reproducing all on--shell correlation functions at tree level, and
next we constructed the unique (up to total derivatives)
spacetime action for the K\"ahler potential, which reduced to the
Ooguri--Vafa action when expanded around Minkowski space.

Additional terms with vanishing on--shell contributions could also
have been added; however, \eSTOT\ is the unique {\it minimal} action
reproducing all on--shell amplitudes.  Its invariance under the same
off--shell symmetry algebra found in section 3 in the
decompactification limit, sdiff$_k(M)$, provides strong evidence that
this is the correct off--shell extension of the on--shell algebras of
area--preserving diffeomorphisms of each of the null planes.  This
infinite--dimensional gauge symmetry will strongly constrain the form
of higher--loop corrections.

\newsec{Discussion: The Heterotic ($N=2$) String}
We have studied the underlying gauge symmetry structure of the closed
$N\nobreak =\nobreak 2$ superstring in toroidal backgrounds.  Within
the natural underlying symmetry algebra vdiff$(T^{4,4})$, the
contribution of the left-moving and the right-moving sectors of the
closed string are manifest. Moreover, the underlying algebra is
background--independent, as all $(4,4)$--dimensional Narain tori are
isomorphic.

The next step towards revealing the underlying structure of the $N=1$
heterotic string is to determine the underlying symmetries of the
$N=2$ heterotic string. This is left for future work; here we will
limit ourselves to a few speculations.

The left-handed sector of the critical $N=(2,0)$ heterotic string is
the $N=2$ superstring in spacetime signature (2,2), and the
right-handed sector is a bosonic string in $(2,26)$ dimensions
\ref\rOV2{H.  Ooguri and C. Vafa, Nucl. Phys.  {\bf B367} (1991) 83.}.
To find a maximal on--shell gauge algebra, we may consider total
compactification on a (4,28)--dimensional Narain torus of the form
$T^{2,2}_L\times T^{2,26}_R$. As before, the left--moving on--shell
algebra will generate area--preserving diffeomorphisms of a null
2-torus, while the right--moving on--shell algebra will include both
gauge and diffeomorphism generators. As a maximal on--shell sublattice
of $\Gamma^{2,26}$ we may take the lattice generated by the root vectors
of $g=E_8^3$ (or any rank 24 Niemeir lattice)
and 2 perpendicular null vectors. If
there had just been one perpendicular null vector, the resulting
algebra of currents would  have been the affine Lie algebra $\hat
g$ \rGP\ref\rKAC{V. Kac, ``Infinite Dimensional Lie Algebras," 2nd ed.
(Cambridge: Cambridge Univ. Press, 1985).}. But with two null
vectors available, we have a doubly--indexed affine Lie algebra.  Now
as before, there are many choices for on--shell algebras, and there
will always be points in the moduli space of (4,28) Narain lattices at
which any given on--shell algebra is realized exactly.  The off--shell
algebra must include all possible on--shell algebras as subalgebras,
and in particular, symplectic diffeomorphisms associated to all null
momenta in $\Gamma^{4,28}$ must be included. Thus it seems reasonable
to expect that the full off--shell algebra will contain vdiff$(T^{4,28})$.
This, however, can not be the full algebra, which in addition must
include generators corresponding to $e^{ip_Rx_R}$,
where $p_R$ are length 2 vectors
of the rank 24  internal lattice.

It is also plausible that the underlying symmetry structure of the
$N=1$ heterotic string is related to
vdiff$(T^{10,26})$,  $T^{10,26}$ being the Narain torus of the
totally compactified target space. However, the proliferation of
oscillator modes leads us to suspect that in this case the algebra
will be much larger, perhaps as large as the loop algebra ${\cal
L}({\rm vdiff}(T^{10,26}))$.

\bigskip
{\bf Acknowledgments}: We thank B.\ Greene, A.\ leClair, R.\ Plesser,
M.\ Ro\v cek, D.\ Spector, H.\ Tye, E.\ Verlinde, E.\ Witten, and B.\
Zwiebach for discussions.  The work of AG is supported in part by DOE
grant no.  DE-FG02-90ER40542. AS is supported in part by NSF grant no.
PHY--87--15272.

\listrefs
\end